\documentclass[aps,twocolumn,amsmath,amssymb,showpacs,prl,superscriptaddress,unsortedaddress]{revtex4}
\usepackage{epsf}
\usepackage{graphicx}
\usepackage{sidecap}

\newcommand{\etal}{{\it et al.}}

\begin{document}

\title{Fermi surface topology and low-lying electronic structure of 
a new iron-based
superconductor Ca$_{10}$(Pt$_3$As$_8$)(Fe$_2$As$_2$)$_5$}

\author{M.~Neupane\footnote{These authors contributed equally to this work.}}
\affiliation {Joseph Henry Laboratory and Department of Physics,
Princeton University, Princeton, New Jersey 08544, USA}

\author{Chang~Liu$^*$}
\affiliation {Joseph Henry Laboratory and Department of Physics,
Princeton University, Princeton, New Jersey 08544, USA}

\author{S.-Y.~Xu}
\affiliation {Joseph Henry Laboratory and Department of Physics,
Princeton University, Princeton, New Jersey 08544, USA}

\author{Y.~J.~Wang}
\affiliation {Department of Physics, Northeastern University,
Boston, Massachusetts 02115, USA}

\author{N.~Ni}
\affiliation {Department of Chemistry, Princeton University,
Princeton, New Jersey 08544, USA}

\author{J. M. Allred}
\affiliation {Department of Chemistry, Princeton University,
Princeton, New Jersey 08544, USA}

\author{L.~A.~Wray}
\affiliation{Joseph Henry Laboratory and Department of Physics,
Princeton University, Princeton, New Jersey 08544, USA}
\affiliation{Advanced Light Source, Lawrence Berkeley National
Laboratory, Berkeley, California 94305, USA}

\author{H.~Lin}
\affiliation {Department of Physics, Northeastern University,
Boston, Massachusetts 02115, USA}

\author{R.~S.~Markiewicz}
\affiliation {Department of Physics, Northeastern University,
Boston, Massachusetts 02115, USA}

\author{A.~Bansil}
\affiliation {Department of Physics, Northeastern University,
Boston, Massachusetts 02115, USA}

\author{R.~J.~Cava}
\affiliation {Department of Chemistry, Princeton University,
Princeton, New Jersey 08544, USA}

\author{M.~Z.~Hasan}
\affiliation {Joseph Henry Laboratory and Department of Physics,
Princeton University, Princeton, New Jersey 08544, USA}

\date{\today}
\begin{abstract}

We report a first study of low energy electronic structure and Fermi
surface topology for the recently discovered iron-based
superconductor Ca$_{10}$(Pt$_3$As$_8$)(Fe$_2$As$_2$)$_5$ (the 10-3-8
phase, with $T_c \sim 8$ K), via angle resolved photoemission
spectroscopy (ARPES). Despite its triclinic crystal structure, ARPES
results reveal a fourfold symmetric band structure with the absence
of Dirac-cone-like Fermi dots (related to magnetism) found around the Brillouin zone
corners in other iron-based superconductors. Considering that the
triclinic lattice and structural supercell arising from the
Pt$_3$As$_8$ intermediary layers, these results indicate that those
layers couple only weakly to the FeAs layers in this new
superconductor at least near the surface, which has implications for the determination of its
potentially novel pairing mechanism.

\end{abstract}


\maketitle

The recent discovery and characterization of new superconducting
phases in the Ca-Fe-Pt-As system
Ca$_{10}$(Pt$_n$As$_8$)(Fe$_2$As$_2$)$_5$ \cite{Kudo, Ni, Car, Kak}
has potentially significant impact on the field of iron-based
high-$T_c$ superconductors \cite{Kamihara, Rotter, Johnston,
Canfield, Canfield_Review}. Most importantly, these new phases serve
as ideal platforms for systematic studies of the physics of the
intermediary layers and their impact on the superconducting
properties - an important yet open question in the field of arsenide
superconductivity. In high-$T_c$ cuprates \cite{Bednorz, Cu_book1,
Cu_book2, cuprate, cuprate2}, such a study is made possible by the
availability of e.g. the Bi$_2$Sr$_2$Ca$_{n-1}$Cu$_n$O$_{2n+4+x}$
($n$ = 1 - 3) series \cite{cuprate, cuprate2}, within which one
finds a drastic correlation between the number of intermediary
layers and the superconducting transition temperature ($T_c$). In
the iron pnictides, a similar type of survey has previously been
unavailable due to the lack of appropriate systems: one needs to
search for a series of stoichiometric materials with different but
systematically adjusted chemical compositions in either the
iron-containing layers or the intermediary layers. The unique
crystal structures in the Ca-Fe-Pt-As systems, on the other hand,
yield drastically different symmetries and periodicities for the
layers of FeAs tetrahedra and the intermediary layers. Therefore,
the intralayer (hopping within the FeAs layers) and interlayer
(hopping between the FeAs and Ca-Pt-As layers) contribution to the
density of states at the Fermi level ($E_F$) can be uniquely
distinguished. Studies of the electronic structure of the
Ca-Fe-Pt-As system are thus of crucial importance toward the
understanding of the interlayer physics and the microscopic
mechanism of high-$T_c$ superconductivity in the iron pnictides.

\begin{figure}[b]
\centering
\includegraphics[width=9cm]{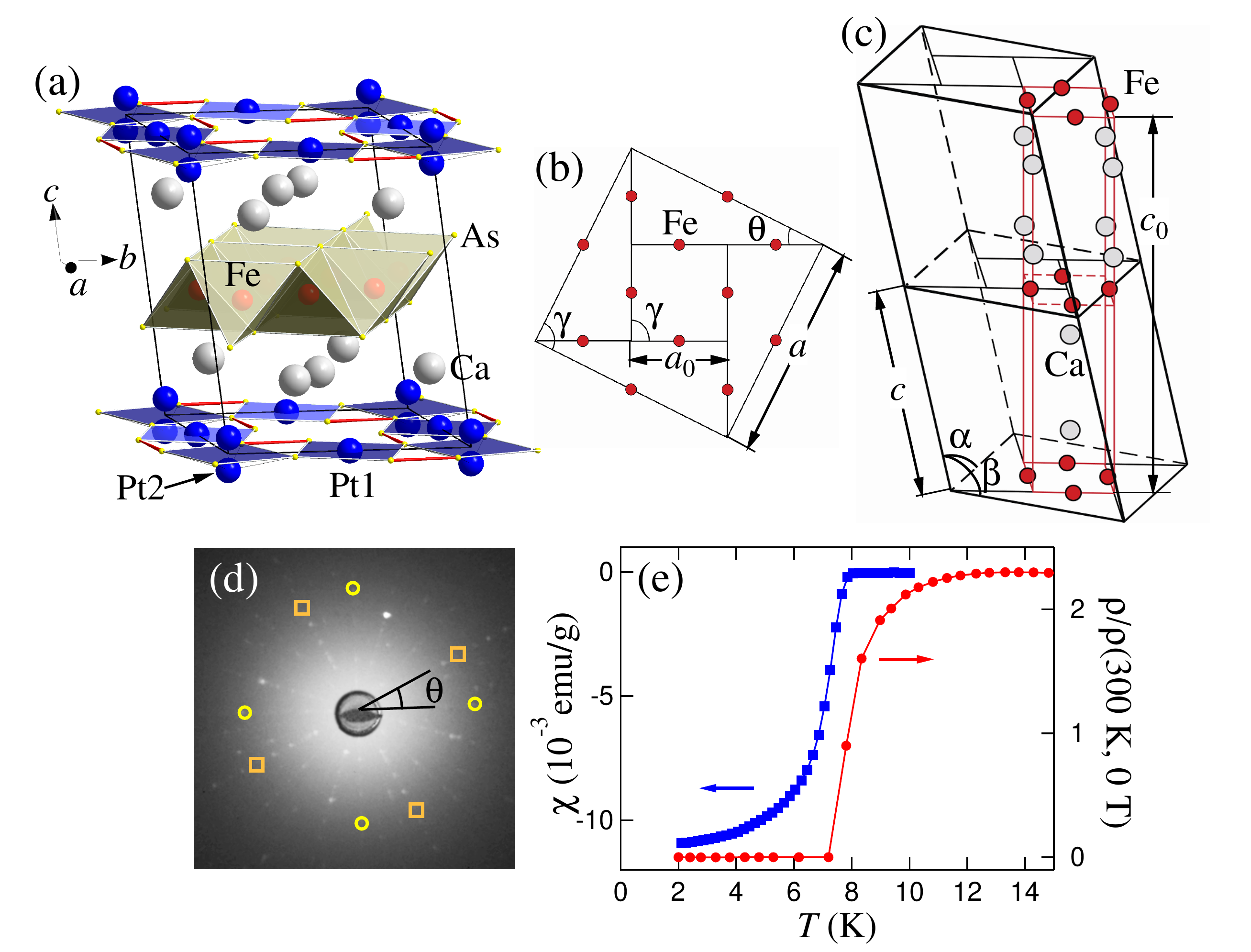}
\caption{Crystallographic and superconducting properties of Ca$_{10}$(Pt$_3$As$_8$)(Fe$_2$As$_2$)$_5$ (10-3-8). (a) Crystal structure of the 10-3-8 phase. (b)-(c) Schematic crystal structure
in (b) the $a$-$b$ plane and (c) three dimensions. (d) Laue picture
of a 10-3-8 single crystal. Reflection peaks for both the triclinic
(yellow circles) and tetragonal (orange squares) lattices are
clearly observed. (e) Low temperature resistivity (right axis) and
magnetic susceptibility (left axis) for the 10-3-8 phase.}
\end{figure}

\begin{figure*}
\centering
\includegraphics[width=17cm]{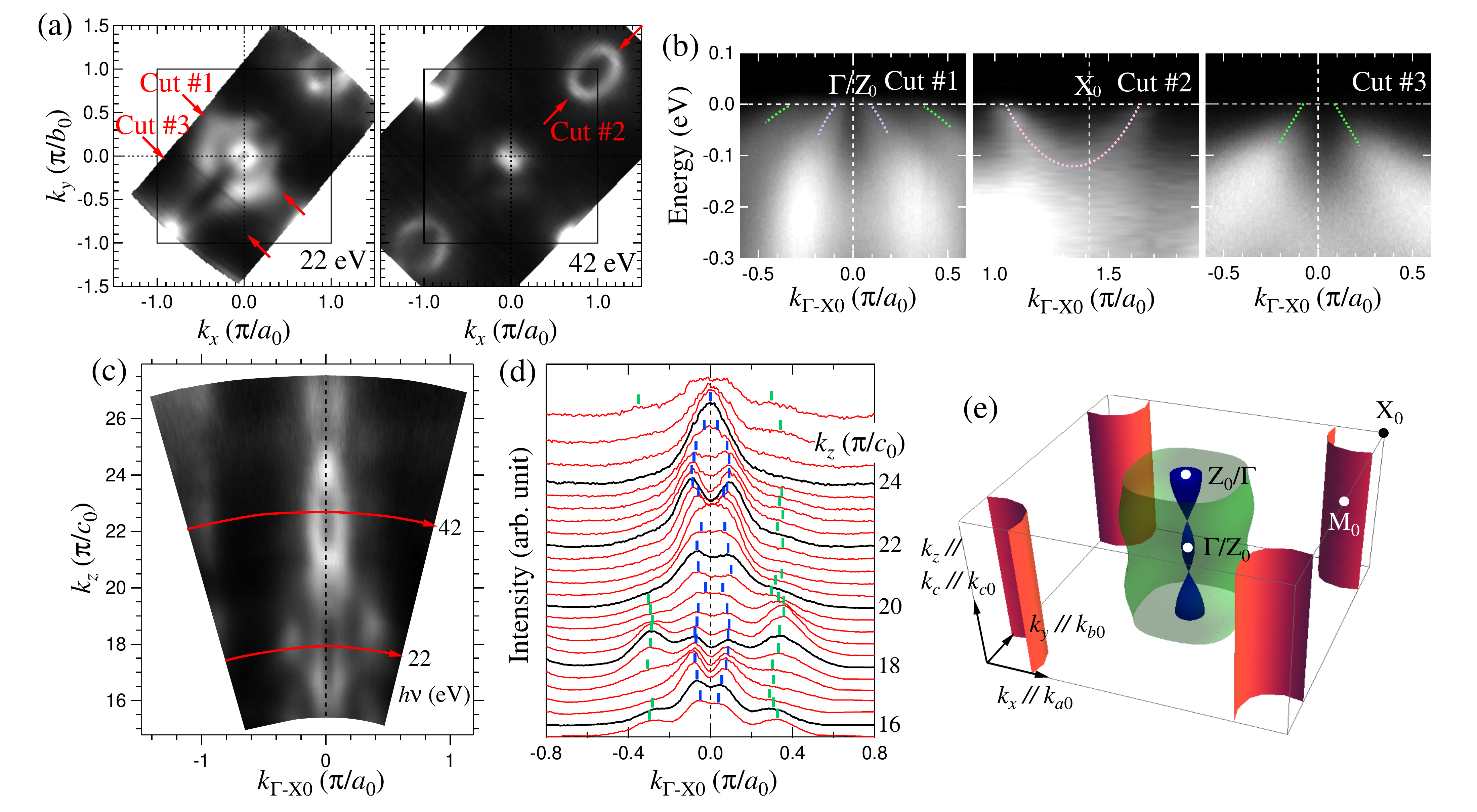}
\caption{Three dimensional ($k_z$ dispersion) ARPES Fermi surface topology. (a) Fermi
surface maps for two different photon energies (lower right corner
in each panel). Brillouin zone sizes are determined based on the
tetragonal cell [see Fig. 2(e) for relations between momenta in the
triclinic and tetragonal cells]. (b) Raw $k$-$E$ maps along
directions marked by Cut \#1 - \#3 in Fig. 2(a). The two hole
pockets around $\Gamma/Z_0$ and the electron pocket around $X_0$ are
labeled as $\alpha_1$, $\alpha_2$ and $\beta_1$, respectively and
are tracked with blue, green and pink colors. (c) $k_z$ dispersion
data for the 10-3-8 phase, taken along the $\Gamma$-$X_0$ direction
with photon energies 15 to 64 eV. Inner potential is set to be 9.5
eV. (d) Momentum distribution curves (MDCs) for different $k_z$
values. Fermi crossing bands are marked with the same colors as in
Fig. 2(b). (e) Schematic experimentally-derived Fermi surface
construction in three dimensions.}
\end{figure*}

In this paper, we report a study of the electronic structure of the
Ca-Fe-Pt-As system for $n$ = 3 [the 10-3-8 phase with $T_c \sim 8$
K, see Fig. 1(e)] using angle resolved photoemission spectroscopy
(ARPES) as well as first principle calculations. ARPES measurements
reveal the three dimensional Fermi surface topology and the band
structure close to $E_F$. We observe well defined Fermi surfaces
with \textit{tetragonal} symmetry that are similar to other
iron-based superconductors, even though arising from an
unambiguously triclinic crystal structure with a larger in-plane
unit cell. First principle band calculations find very small
contribution of the platinum density of states at $E_F$ for the
10-3-8 phase. These results are indicative of a weak interlayer
hopping between the FeAs and the PtAs intermediary layers in this
Ca-Fe-Pt-As system. ARPES data also shows that the quasi nesting
condition between the Fermi pockets at the Brillouin zone (BZ)
center and the BZ corner is not perfect. This may be another reason
for the low $T_c$ of the 10-3-8 phase.

\begin{SCfigure*}
\centering
\includegraphics[width=12.5cm]{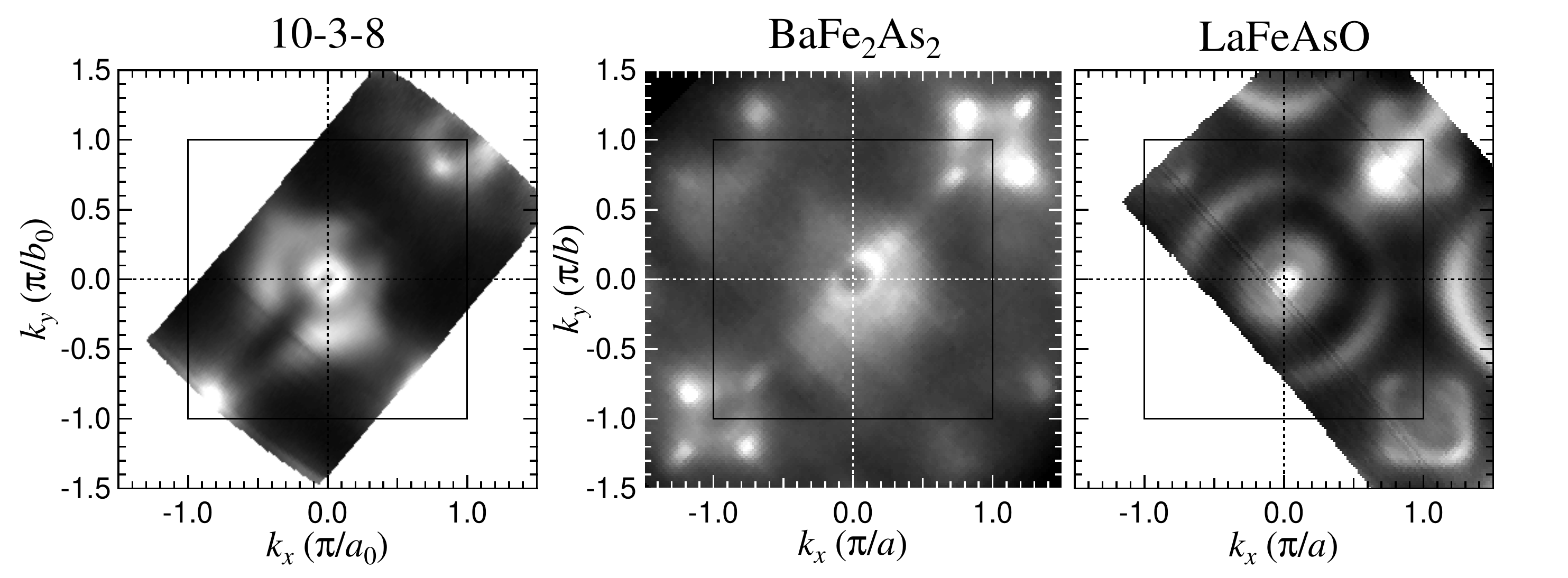}
\caption{Comparison of in-plane ARPES Fermi surfaces of the 10-3-8
phase, BaFe$_2$As$_2$ \cite{Chang_NPhys} and LaFeAsO
\cite{Chang_1111}. Incident photon energies are 22, 105 and 45 eV,
respectively. The Fermi ``dots'' seen around the $X$ pockets of
BaFe$_2$As$_2$ come from antiferromagnetic reconstructions of the
electronic structure \cite{Chang_NPhys}, and the large $\Gamma$ hole
pocket in LaFeAsO originates from surface effects
\cite{Chang_1111}.}
\end{SCfigure*}

We begin our discussion with a detailed examination of the
crystallographic properties of the 10-3-8 single crystal, as
summarized in Fig. 1. This crystal has a triclinic unit cell with
primitive vectors of length $a = b = 8.759$ $\mathrm{\AA}$, $c =
10.641$ $\mathrm{\AA}$, $\alpha = 94.744^\circ$, $\beta =
104.335^\circ$, and $\gamma = 90.044^\circ \simeq 90^\circ$
\cite{Ni} [see Fig. 1(a)-(c) for definitions]. Markedly, such a
triclinic atomic arrangement is experienced only by the platinum
atoms in the crystal; the calcium atoms and the FeAs$_4$ tetrahedra
arrange in a tetragonal fashion with lattice parameters $a_0 = b_0 =
3.917$ $\mathrm{\AA}$, $c_0 = 20.548$ $\mathrm{\AA}$. From Fig. 1(b)
one notices that the in-plane triclinic unit cell is essentially a
$\sqrt{5}$ superlattice that forms along the (210) direction of the
tetragonal cell, making an in-plane inclusion angle of $\theta =
\mathrm{arctan}(1/2) = 26.56^\circ$ and a length relation $a =
\sqrt{5}a_0$ between the two lattices. Fig. 1(c) reveals that the
top and bottom surface centers of the triclinic unit cell displace
horizontally via an in-plane vector ($a_0$/2, $a_0$/2). Therefore
the height of the tetragonal cell measures $c_0 = 2\sqrt{c^2 -
a_0^2/2} = 20.548$ $\mathrm{\AA}$. Hence, if the interlayer hopping
between the PtAs and FeAs layers is weak, the ARPES Fermi surface
should reflect a similar topology to the other prototype pnictides,
with possibly weak FeAs shadow bands associated with superlattice
folding and additional features associated with the PtAs layer. It
is important to point out that any potential crystal twinning would
not hinder the ARPES observation of the PtAs superlattice. See
online supplementary information (SI) for the detailed description
of Fig. 1(d). From the present experiments, we cannot rule out the
possibility that subtle surface reconstruction gives rise to a
surface-driven electronic structure which has higher symmetry than
that of the bulk.

We now present the ARPES data for the Fermi surface topology of the
10-3-8 phase. Fig. 2 shows the three dimensional electronic
structure obtained by ARPES and a schematic three dimensional Fermi
surface structure deduced from our experiments (see SI for a
detailed description). The most important observation from Fig. 2 is
that the ARPES electronic structure has a \textit{tetragonal}
symmetry, and the experimental Brillouin zone size is proportional
to $\pi/a_0$ rather than $\pi/a$ in the $k_x$-$k_y$ plane [Fig.
2(a)]. In other words, the ARPES signal reveals that the electronic
system is tetragonal at least near the surface, with the periodicity of the FeAs layer
sublattice. This observation is noteworthy for two reasons. First,
it points out directly that the triclinic arrangement and larger
supercell periodicity of the platinum atoms has very little
influence on the electronic structure; if the platinum orbitals had
a strong contribution at $E_F$, then the observation of Fermi
pockets arranged according to the triclinic Brillouin zone is
expected - similar to what is shown in Fig. 4(c). From this we
deduce that the hybridization between bands from the PtAs
intermediary layers and those from the FeAs layers has to be weak.
Although this is only a qualitative statement, the unique crystal
structure of the 10-3-8 phase does provide an important estimation
of the interlayer hopping strength that is otherwise hard to obtain
from experiment in the Fe-As superconductors: interlayer hopping
must be so weak that it renders the triclinic lattice invisible by
photoemission. Second, the ARPES electronic structure mimics the
electronic structures of other prototype pnictides like
$AE$Fe$_2$As$_2$ (``122'', $AE$ = Ca, Sr, Ba, etc.). Not only the
in-plane lattice parameter $a_0$ but also the shapes, sizes and
Fermi velocities of the $\Gamma$ and $X_0$ Fermi pockets show very
little difference with those for the 122 parent compounds
\cite{David, Wray, Zabolotnyy, Chang_3D, Kondo, Ding2011} (see SI). This
indicates that a universal electronic structure capturing the
underlying superconducting mechanism may exist for different
sub-families of the Fe-based superconductors (except for the
K$_\alpha$Fe$_{2-\beta}$Se$_2$ series \cite{Zhang_NMat, Qian_Se}).

Despite the overall similarity, there are observable differences
between the Fermi surface of the 10-3-8 phase and that of the
prototype pnictides. In Fig. 3 we compare explicitly the in-plane
ARPES Fermi surfaces for the 10-3-8 phase, BaFe$_2$As$_2$
\cite{Chang_NPhys} and LaFeAsO \cite{Chang_1111}. First, the
Dirac-cone-like Fermi dots around the $X$ points in BaFe$_2$As$_2$
are absent in the 10-3-8 phase [seen most clearly in the 42 eV panel
of Fig. 2(a)]. Since these dots are direct consequences of the long
range antiferromagnetic order present in the 122 compound
\cite{Chang_NPhys, Zahid}, their absence is consistent with the
absence of an antiferromagnetic signature in transport measurements
up to room temperature \cite{Ni}. Second, extended ARPES intensity
along one of the $\Gamma$-$X_0$ directions is seen only for the
10-3-8 phase. A close look into the $k$-$E$ maps (data of Cut \#3 in
Fig. 2) reveals that there is an actual Fermi crossing in these
locations, and that the band is holelike. We point out here that
this band also originates from the Fe orbitals, since it shows
reflective symmetry along the high symmetry directions of the iron
unit cell rather than the triclinic platinum cell. Third, from the
$k_z$ dispersion data [Fig. 2(c)-(d)], one notices that ellipsoidal
hole pockets are observed around both the $\Gamma$ and $Z_0$ points
for the 10-3-8 phase, while in the 122 parent compounds only $Z$
ellipsoids are observed \cite{Chang_3D, Kondo}. According to band
calculation (Fig. 4), this signifies the weak but existent Pt
influence on the Fermi surface topology. The nesting condition for
the $\Gamma/Z_0$ hole pockets and the $X$ electron pockets is
reasonable but not perfect; this may as well be a reason for the
relatively low $T_c$ detected.

\begin{figure}
\centering
\includegraphics[width=9cm]{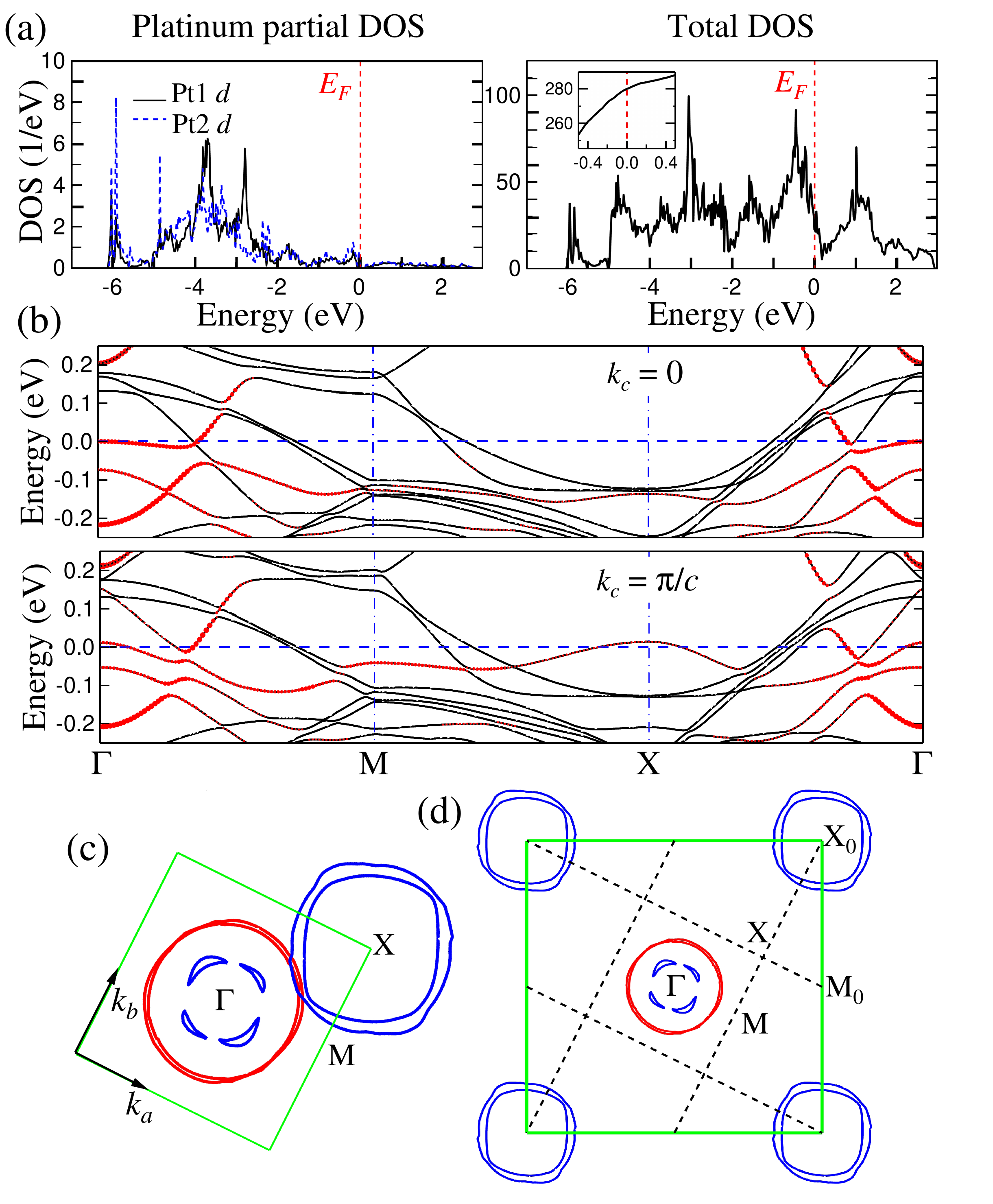}
\caption{Results of first principle calculations. (a) Calculated
density of states (DOS) for the 10-3-8 phase. Left column: partial
DOS for in-plane and out-of-plane platinum $5d$ orbitals (Pt1 \&
Pt2, respectively). Right column: Total DOS for all elements. Inset
shows the integrated DOS (IDOS) near $E_F$. (b) Calculated band
structure of 10-3-8 in the triclinic Brillouin zone. Red dots
indicate the contribution of platinum orbitals. (c) Fermi surfaces
sketched in the triclinic zone. (d) Fermi surfaces unfolded into the
tetragonal zone.}
\end{figure}

We now examine the electronic structure of the Ca-Fe-Pt-As system
from results of first principle calculations \cite{xc, paw, vasp}.
In the calculation, spin-orbit coupling effects are not included and
$E_F$ is fixed assuming zero electron doping. The crystal structure
of the 10-3-8 phase is taken from Ref. \cite{Ni} with the off-plane
Pt (Pt2) atom in the (Pt$_{3}$As$_{8}$) layer assigned to one of its
two possible positions - slightly above the plane \cite{Ni} (see
also SI). We show in Fig. 4(a) the calculated curves for the total
density of states (DOS) as well as the partial DOS (PDOS) for the
two different kinds of platinum atoms [Pt1 and Pt2, see Fig. 1(a)].
Similar to the prototype pnictides, the DOS in vicinity of $E_F$ for
the 10-3-8 phase is dominated by Fe 3$d$ orbitals. In Fig. 4(b)-(c),
the calculated LDA band structure and Fermi surface are shown in the
triclinic Brillouin zone. From Fig. 4(b) we see that the innermost
$\Gamma$ band has stronger $k_z$ dispersion than the other two
holelike iron bands, which is consistent with the ARPES observations
[Fig. 2(c)-(e)]. This band is hybridized with the Pt $d_{xz}$
orbitals. Since the experimental data reveal no sign of the
triclinic Fermi surfaces, we assume that the potential from the
$\sqrt{5}$ superlattice arising from the Pt$_{3}$As$_{8}$ layers is
very weak, in which case we can approximately ``unfold'' the LDA
superlattice band structure into the tetragonal BZ, as shown in Fig.
4(d). The tetragonal $\Gamma$, $X_0$ and $M_0$ points form a subset
of the supercell $\Gamma$, $X$ and $M$ points, so we systematically
``erase'' all superlattice FS maps not associated with the
tetragonal symmetry points. From Fig. 4(d), we see that this process
works well for the Fe Fermi surfaces, reproducing the familiar 122
structure. The fate of the electronlike Pt-bands (the four
pedal-like small pockets near $\Gamma$) under this unfolding process
is less clear, but it should be noted that these pockets are not
clearly distinguished by ARPES.

Interlayer hopping plays an important role in the presence of
high-$T_c$ superconductivity in both the cuprates and the Fe-based
superconductors. It is believed that the hopping strength in the
prototype pnictides is somewhat stronger than that in the cuprates
(although still on the weak side), and that this may lead to the
differences between $T_c$ and pairing symmetry in these two
families. Zhai \textit{et al.} propose that the superconducting gap
symmetry changes from $d$-wave to $s$-wave with increasing hopping
strength \cite{Zhai}. Although the model in Ref. \cite{Zhai} may not
apply directly to the Ca-Fe-Pt-As systems, where spin density wave
signatures have not yet been detected at low temperatures, it is
likely that the unique momentum-space separation of the inter- and
intra-layer signals in the 10-3-8 system helps to determine the
hopping strength with considerably higher accuracy, thus shedding
light on the ultimate determination of the superconducting mechanism
in both classes of high-$T_c$ superconductors.

In conclusion, we have presented a systematic study of the band
structure and Fermi surfaces of one of the new Ca-Fe-Pt-As
superconductors in the vicinity of $E_F$. Our ARPES observation -
reduced tetragonal electronic structure and little $k_z$ dispersion
- points to a weak interlayer hopping strength in this system. The
Dirac-cone-like Fermi dots around $X$ are absent in the 10-3-8
phase, consistent with the absence of long range antiferromagnetism
in this compound. The nesting condition between the Fermi pockets at
the center and the corners of the Brillouin zone is not perfect.
This may contribute to the low $T_c$ of the 10-3-8 phase. First
principle calculations agree well with experimental data if the
potential from the $\sqrt{5}$ superlattice arising from the PtAs
layers is considered to be very weak, and the triclinic band
structure can be unfolded onto the tetragonal Brillouin zone. The
Ca-Fe-Pt-As superconductors are an ideal system for the study of
interlayer hopping in the iron-based superconductors using many different techniques. The present
detailed ARPES study of the electronic structure of the 10-3-8 phase near the surface
serves as an essential first step in that direction.

 Work at Princeton is supported by NSF-DMR-1006492
and the AFOSR MURI on superconductivity. The Synchrotron Radiation
Center is supported by NSF-DMR-0537588. C. L. acknowledges T. Kondo and A. Kaminski for provision of data
analysis software. M. Z. H. acknowledges
Visiting Scientist support from LBNL and additional support from the
A. P. Sloan Foundation.

\end{document}